# MULTIMODAL REGISTRATION OF THE FACE FOR COMPUTER-AIDED MAXILLOFACIAL SURGERY

TH. LELOUP (1), M. CHABANAS (2), Y. PAYAN (2)

(1) Information and Decision System Dept – Univ. of Brussels
50, Av. Roosevelt CP165/57 – 1050 Brussels – Belgium
(2) TIMC-GMCAO Laboratory – IN3S – Faculté de Médecine
38706 La Tronche – France

## INTRODUCTION

Maxillofacial dysmorphosis of the lower part of the face (disequilibrium between the mandible, the upper jaw and the face) has important functional, orthodontic and aesthetic consequences : respiration difficulties, mastication and elocution troubles, disruptions of the dental occlusion, face asymmetry… Orthognathic surgery can correct these problems but is very delicate. It mainly consists in osteotomies and bone segment repositioning in order to realign the upper and lower jaws (Fig. 1). This surgery requires precise planning of bone structure displacement. One of the main request of the patients concerns the prediction of the face aesthetic after the operation.

A complete protocol for computer-aided maxillofacial surgery was already presented [1]. This protocol includes several important steps :
− Simulation of bone osteotomies on a virtual 3D model of the patient skull,
− Planning of the bone segment repositioning, with six degrees of freedom, using 3D cephalometric analysis,
− Quantitative measurement of the dental occlusion,
− Prediction of the patient facial soft tissues deformations,

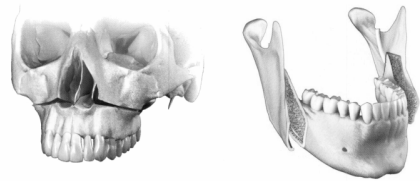

*Fig. 1 : Maxillar and mandibular osteotomies*

− Computer-aided intervention in the operating room thanks to the 3D bone repositioning panning.

To predict the soft tissues modifications resulting from the repositioning of the underlying bone structures, a generic finite element model of the face soft tissues is adapted to the patient by the Octree Spline elastic registration method [6]. The skin and skull surfaces of the patient are segmented from the CT-scanner exam. The generic model is composed of several types of nodes : external nodes modelling the skin, intermediate nodes delimiting the dermis and the hypodermis and internal nodes mainly located on the skull (Fig. 2). The initial position of the generic model is determined manually. Both generic model and patient surfaces are considered as clouds of points by the registration process. A first elastic registration is effected to match external nodes of the generic mesh to the patient skin surface. The obtained transformation is applied to all nodes of the generic model. A second elastic registration is computed to match the internal nodes in contact with the skull to the patient skull surface. At this stage, some manual adaptation can be effected if the generic model do not match exactly the patient one. Moreover, the transformed generic mesh is corrected in order to fulfill regularity criteria needed for finite element computation.

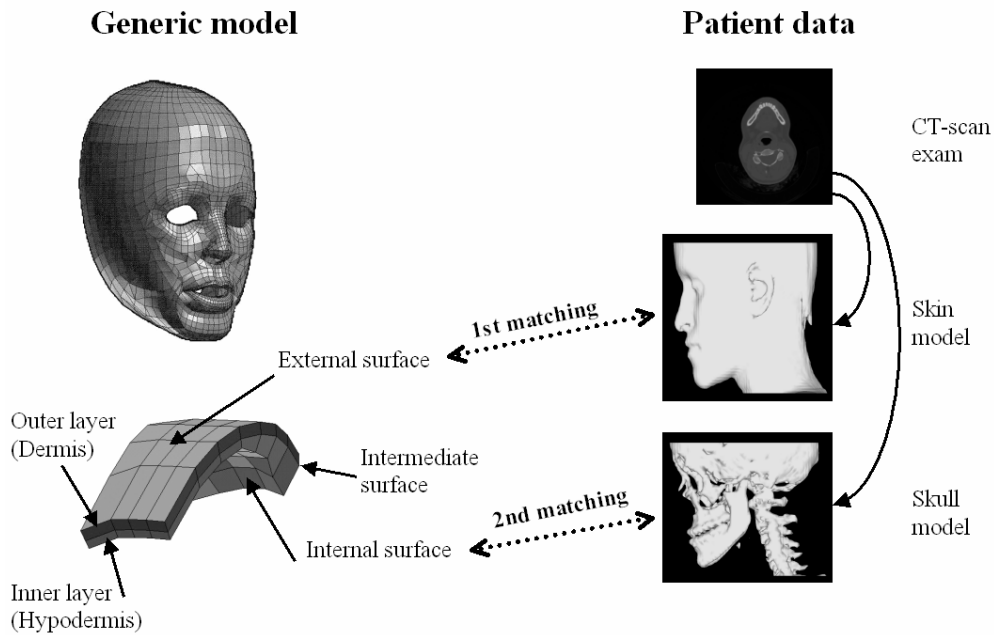

*Fig. 2 : Adaptation of the finite element generic model to the patient*

## OBJECTIVES

The generic face model adaptation method described above works pretty well in many cases but several disadvantages have been identified :
- The initial position of the generic model is performed by manual interaction, which can lead to significantly different registration results, according to the skill of the user.
- The patient skin surface obtained from the CT-scanner exam is built by the Marching Cube algorithm and includes consequently internal structures of the skin (sinus, trachea…). This can disturb in many cases the registration process (the skin surface of the generic model is often matched on the sinus surface of the patient model, for example). The same problem occurs also with the skull surface. This is due to the point-to-point registration process that does not consider the models as "real surfaces" but only as clouds of points.

This paper present a new approach to perform accurate automatic face registration, based on multimodal data including cephalometric analysis and surface information.

## METHODS

The first objective is to determine automatically a initial position for the generic model. To perform this task, it is a good idea to use the cephalometric data. They consist in a series of particular anatomic points, easily identifiable and defined in 3D on the CT-scanner exam of the patient. The surgeon can perform cephalometric analysis [2,4], based on planes, angles, surfaces or volumes defined from these anatomic points (Fig. 3). For the generic model, these points are defined once for all. It implies that two clouds of corresponding points are available in patient and generic models.

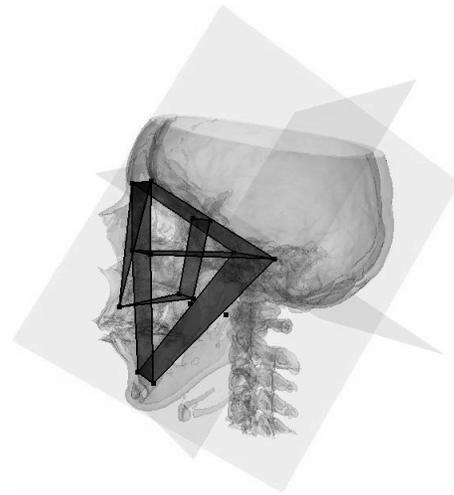

*Fig. 3 : Three-dimensional cephalometric analysis based on a series of anatomic points*

Our first idea was to compute an initial position by rigid registration of these corresponding points. However, patients suffering from maxillofacial dismorphosis present very different skull morphologies and even after this step the generic model can be located relatively far from the patient one. Therefore, we have preferred to use an elastic registration to match the two clouds of corresponding points. This do not need much more time and the resulting transformed generic model corresponds better to the patient (Fig. 4).

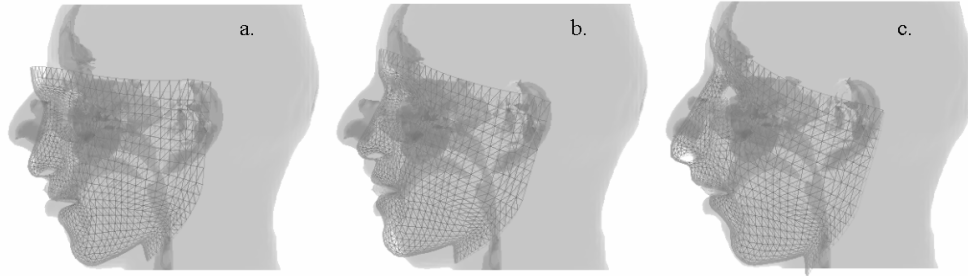

*Fig. 4 : Automatic determination of the initial position based on feature points – a. Original position (scanner dependant); b. Rigid registration; c. Elastic registration. The generic model is in wireframe.*

The second objective concerns the integration of surface information into the elastic registration process. The Octree Spline elastic registration algorithm compute a transformation that minimizes an energy function which represents a weighted sum of distances squares. To calculate these distances, a corresponding point located on the target model (patient) has to be determined for each point of the source model (generic mesh). The point-to-point registration process searches only for corresponding points among nodes of the target model. Two improvements can be added to this method.

The first one consists in taking into account the relative local orientation of the surfaces to match [3,5]. For each model node, the external normal can be computed thanks to the adjacent triangles. Then the search for a corresponding target point can be effected according to a double criterion : minimum Euclidian distance and near normal orientation (i.e. normal scalar product near to one). The corresponding point is searched only in an isotropic neighbourhood centred on the considered source point. A new composed distance $Dist_{Comp}(P, Q)$ is therefore defined between two points P and Q :

$$Dist_{Comp}(P, Q) = Dist_{Eucl}(P, Q) + w \cdot R \cdot (1 - N_P \cdot N_Q)/2$$

$Dist_{Eucl}$ represents the standard Euclidian distance. $N_P$ and $N_Q$ are the normals associated to the points P and Q; R is the radius of the isotropic neighbourhood and w is the weighting factor of the normal orientation term. Note that the scalar product is rescaled between 0 and 1 in order to be high when orientations are discordant and null when normals are oriented identically.

The orientation of both source and target models surfaces is thus taken into account (Fig. 5), which allows to match the surfaces more efficiently : source points are coupled with target points that correspond better anatomically.

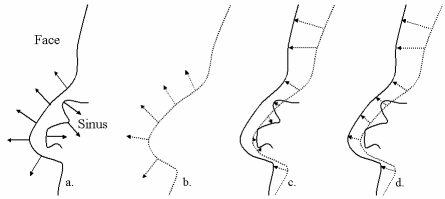

*Fig. 5 : Utility of using surface normals – a. Target model; b. Source model; c. Matching by smallest Euclidian distance; d. Using distances and normals.*

The second improvement is to search corresponding target points on the whole target surface and not only among target model nodes. This task is done in two steps : in a first time, the corresponding target node is determined thanks to the composed distance; in a second time, the closest corresponding point (using the composed distance) is searched in all the adjacent triangles of the previous node by recursive subdivision of each triangle.

## RESULTS

Our method has been tested and validated on a set of 10 patients faces.

The improvements of our work have been quantified by measuring statistical distance parameters (maximum, mean, median, standard deviation and $95^{th}$ percentile) for both methods : point-to-point matching (PPM) using only distance criterion and point-to-surface matching (PSM) using normals information besides.

Subdivision level of the octree was limited to 5 because higher levels require too many computing resources. However, the matching result can be still improved after only one process. Therefore, two successive registration processes have been effected. The profit becomes negligible afterwards.

Several zones of the generic model are not present on the patient data, depending on how the CT-scanner was done (face, neck, end of the nose…). In the same way, ears are not represented in our generic model but appear in the patient model. These zones, where no matching is possible for lack of data, have been removed manually and are not taken into account in the statistical results (Fig. 6).

Colour distance map is available in order to evaluate globally the matching quality and to localize the possible problem zones (Fig. 7).

We have observed that the matching is particularly difficult in the lips region, where surface orientation can considerably vary from one person to another. This zone often contains the larger errors.

It can also be noticed that with the PSM algorithm, the source model do not remains attached to the internal structures present on the patient model (interior of the nose, sinus…).

Several of our patients have orthodontic braces or teeth fillings which generate CT-scanner image and 3D surface artefacts but our matching technique remains robust.

The time needed to proceed to one matching process is about one minute on Pentium IV 1.6Ghz PC computer, which is clinically acceptable.

| Patient | Matching | Maximum | Mean | Median | Stand Dev | Percentile 95 |
|---|---|---|---|---|---|---|
| 1 | PSM | 2,79 | 0,47 | 0,37 | 0,40 | 1,24 |
| | PPM | 6,40 | 0,51 | 0,36 | 0,61 | 1,38 |
| 2 | PSM | 3,32 | 0,55 | 0,46 | 0,44 | 1,36 |
| | PPM | 3,96 | 0,56 | 0,45 | 0,49 | 1,53 |
| 3 | PSM | 3,57 | 0,52 | 0,39 | 0,47 | 1,41 |
| | PPM | 7,21 | 0,69 | 0,47 | 0,72 | 2,23 |
| 4 | PSM | 1,97 | 0,37 | 0,28 | 0,32 | 0,99 |
| | PPM | 2,74 | 0,42 | 0,29 | 0,39 | 1,23 |
| 5 | PSM | 4,84 | 0,39 | 0,29 | 0,39 | 1,08 |
| | PPM | 2,75 | 0,44 | 0,33 | 0,39 | 1,23 |
| 6 | PSM | 3,93 | 0,41 | 0,30 | 0,42 | 1,07 |
| | PPM | 4,96 | 0,65 | 0,49 | 0,62 | 1,78 |
| 7 | PSM | 3,07 | 0,49 | 0,39 | 0,42 | 1,37 |
| | PPM | 3,33 | 0,54 | 0,40 | 0,49 | 1,54 |
| 8 | PSM | 2,92 | 0,54 | 0,42 | 0,44 | 1,45 |
| | PPM | 4,69 | 0,60 | 0,45 | 0,55 | 1,68 |
| 9 | PSM | 2,92 | 0,51 | 0,43 | 0,41 | 1,33 |
| | PPM | 5,09 | 0,61 | 0,47 | 0,54 | 1,72 |
| 10 | PSM | 2,36 | 0,31 | 0,24 | 0,29 | 0,90 |
| | PPM | 1,93 | 0,36 | 0,28 | 0,32 | 0,99 |
| **Average** | **PSM** | **3,17** | **0,46** | **0,36** | **0,40** | **1,22** |
| | **PPM** | **4,31** | **0,54** | **0,40** | **0,51** | **1,53** |

*Fig. 6 : Quantitative comparison between PSM and PPM algorithms (error distances are in mm)*

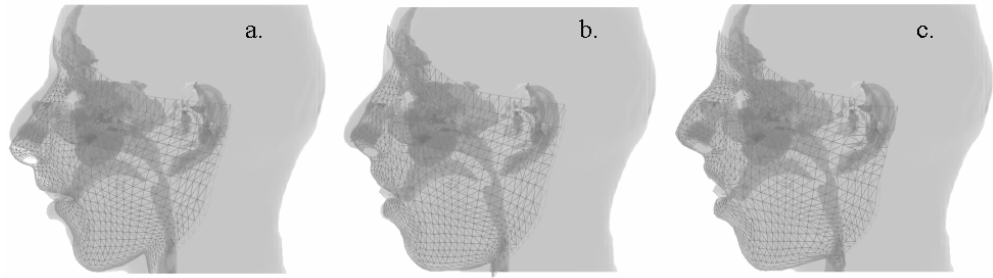

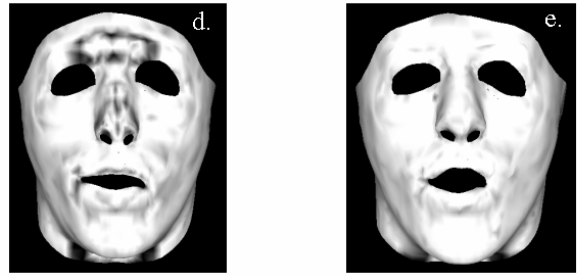

*Fig. 7 : Qualitative comparison between PSM and PPM algorithms – a. Initial situation; b. After two PPM processes; c. After two PSM processes; d. Errors distances map corresponding to the PPM final process; e. Errors distances map corresponding to the PSM final process*

# CONCLUSION

The multimodal elastic registration algorithm presented in this paper has been validated to match a generic model of the face to several patients. This method is automatic, precise and robust. The computing time is compatible with clinical practice constraints.

Future work includes mesh regularity verification in order to insure finite element computation feasibility. It will then be possible to simulate soft tissue deformations resulting from bone repositioning during maxillofacial surgery.

# AKNOWLEDGMENT

This work was supported by the "*Fédération pour la Recherche Médicale (F.R.M.)*".